\documentclass[aps, prc, reprint, amsmath, amssymb, superscriptaddress, floatfix, nofootinbib]{revtex4-1}

\usepackage[pdftex]{graphicx}% Include figure files
\DeclareGraphicsExtensions{.jpg,.png,.pdf}

\usepackage[utf8]{inputenc}
\usepackage{hyperref}
\usepackage{amsmath}
\usepackage{amssymb}
\usepackage{amsfonts}
\usepackage{tabularx}
\usepackage{booktabs}
\usepackage{graphicx}
\usepackage{color}
\usepackage{multirow}
\usepackage[perpage,symbol]{footmisc}
\usepackage{verbatim}
\usepackage{dcolumn}% Align table columns on decimal point
\usepackage{bm}% bold math
\usepackage[inline]{enumitem}
\definecolor{theblue}{RGB}{0,50,230}
\hypersetup{
  colorlinks=true,
  linkcolor=theblue,
  citecolor=theblue,
  urlcolor=theblue
}

\newcommand{\pt}{\ensuremath{p}_{\rm T}}
\newcommand{\raa}{\ensuremath{R}_{\rm AA}}
\newcommand{\vtwo}{\ensuremath{v}_{\rm 2}}
\newcommand{\snn}{\sqrt{s_{\rm NN}}}
\newcommand{\s}{\sqrt{s}}
\begin{document}

%\preprint{APS/123-QED}

\title{Charm-strange meson production in ultra-relativistic heavy-ion collisions at the CERN-LHC energies}% Force line breaks with \\

\author{Shuang~Li}
\email{lish@ctgu.edu.cn}
\affiliation{%
College of Science, China Three Gorges University, Yichang 443002, China\\
}%
\affiliation{%
Key Laboratory of Quark and Lepton Physics (MOE), Central China Normal University, Wuhan 430079, China\\
}%
\author{Chaowen~Wang}%
\email{wangchaowen@ctgu.edu.cn}
\affiliation{%
College of Science, China Three Gorges University, Yichang 443002, China\\
}%

\date{\today}% It is always \today, today, but any date may be explicitly specified

\begin{abstract}
The nuclear modification factor $\raa$ and the elliptic flow coefficient $\vtwo$ of charm-strange meson $D^{+}_{s}$
is systematically studied in Pb--Pb collisions at $\snn=5.02~{\rm TeV}$ and $2.76~{\rm TeV}$.
During the modeling, the coupling strength between the injected charm quark and the incident medium constituents,
is extracted from the lattice QCD calculations:
$2\pi TD_{s}=7$ (\textbf{Model-A}) and $2\pi TD_{s}=1.3 + (T/T_{c})^2$ (\textbf{Model-B}).
We find that, comparing $\raa(D^{+}_{s})$ with $\raa(non-strange)$,
the heavy-light coalescence effect is more pronounced for the former one,
resulting in an enhancement behavior in the range $2\lesssim\pt\lesssim5~{\rm GeV}$.
The predictions for $\raa(D^{+}_{s})$ and $\raa(non-strange)$ favor Model-A to have a better description of the measured $\pt$ dependence in both energies,
while their $\vtwo$ prefer Model-B at moderate $\pt$ ($2\lesssim\pt\lesssim4~{\rm GeV}$).
Therefore, it is necessary to consider the temperature- and/or momentum-dependence of $2\pi TD_{s}$ to
describe simultaneously $\raa(D^{+}_{s})$ and $\vtwo(D^{+}_{s})$ in different centrality classes in Pb--Pb collisions.
\end{abstract}

\maketitle

%\tableofcontents

%%==============================================
%%==============================================
%%==============================================
%%==============================================
\section{INTRODUCTION}\label{sec:Introduction}
Ultra-relativistic heavy-ion collisions provide the unique opportunity to produce
and study the properties of the strongly-interacting matter
within the extreme high temperature and energy density environment,
where a phase transition is expected from the ordinary hadron state to its deconfined constituents,
namely Quark-Gluon Plasma (QGP)~\cite{Gyulassy05,Shuryak05}.
Heavy quarks (HQ) such as charm and bottom are of particular
interest amongst the various probes of the QGP~\cite{RalfSummary16, Jurgen17, HQTransportSummary18}.
Due to the large mass, they are mainly produced at the early stage of the collisions
via the hard scattering process, and subsequently interact with the QGP constituents without affecting their mass,
resulting in the negligible re-generation propagating through the medium.
Meanwhile, the HQ flavour is conserved during the interaction with QGP constituents,
therefore, the initial produced HQ will experience the full evolution of the hot and dense medium.

While traversing the QGP medium, a heavy quark will interact with the medium constituents
and thus, lose part of its initial energy via both elastic ($2\rightarrow2$, collisional processes~\cite{EnergyLossCollisionalSW07})
and inelastic scatterings ($2\rightarrow2+X$, including gluon radiation~\cite{GluonRadiation94}),
naming the collisional and radiative energy loss, respectively.
The energy loss effect together with the HQ hadronization mechanisms can be investigated
by measuring the nuclear modification factor
\begin{equation}\label{eq:RAA}
\raa(\pt,y)=\frac{{\rm d}^{2}\sigma_{\rm AA}/{\rm d}\pt {\rm d}y}{{\rm d}^{2}\sigma_{\rm pp}/{\rm d}\pt {\rm d}y},
\end{equation}
of the final heavy-flavor productions
such as the open charmed mesons (i.e. D mesons including $D^{0}$, $D^{+}$, $D^{*+}$ and $D^{+}_{s}$\cite{ShuangQM14}),
where, ${\rm d}^{2}\sigma_{\rm AA}/{\rm d}\pt {\rm d}y$ is the $\pt$ and $y$ double-differential production cross section
in nucleus-nucleus collisions, scaled by the number of binary nucleon-nucleon collisions;
${\rm d}^{2}\sigma_{\rm pp}/{\rm d}\pt {\rm d}y$ is the double-differential result in nucleon-nucleon collisions.
The deviation of $\raa$ from unity is sensitive to the nuclear effects,
e.g. the initial (anti-)shadowing and the subsequent in-medium energy loss.
In addition, the elliptic flow coefficient
\begin{equation}\label{eq:v2}
\vtwo= \biggr \langle \frac{p_{x}^{2}-p_{y}^{2}}{p_{x}^{2}+p_{y}^{2}} \biggr \rangle,
\end{equation}
allows to describe the anisotropy of the transverse momentum, hence,
$\vtwo$ is sensitive to the EoS and initial conditions in the low $\pt$ region,
and it is also able to reflect path-length-dependence of the energy loss at high $\pt$.
 
Many models were developed~\cite{Akamatsu09, HFModelHee08, CaoPRC13, MCatHQsPRC16, PHSDPRC16, POWLANGJHEP18}
to study the comprehensive sets of the available measurements of non-strange charmed meson, e.g. $D^{0}$, $D^{+}$ and $D^{*+}$.
It was realized~\cite{JFLPRL09, JFLCPL15, Das15, CUJET3JHEP16} that the simultaneous description of their $\raa$ and $\vtwo$ requiring
further understanding of the temperature-dependence of the coupling strength ($2\pi TD_{s}$)
between the injected (heavy) quark and the incident medium constituent.
The charm-strange meson $D^{+}_{s}(c\bar{s})$ production is more interesting with respect to non-strange charmed mesons,
since its valence quark content consists of charm and (anti-)strange quark,
which will couple the well-know strangeness enhancement~\cite{StrangnessPRL82}.
$D^{+}_{s}$ spectra will be therefore affected by both the charm conservation
and the strangeness enhancement effects in heavy-ion collisions.
However, few models~\cite{HFModelHee13} were dedicated to investigate the $D^{+}_{s}$ meson spectra,
as well as its $\raa$ and $\vtwo$ until now.

Based on the previous work, we try to address this question by taking into account the various temperature-dependence
of $2\pi TD_{s}$ which are phenomenologically extracted from the lattice QCD calculation,
and then investigate their effects on the observables ($\raa$ and $\vtwo$),
in particular for the charm-strange meson $D^{+}_{s}$ at the LHC energies.
Meanwhile, as pointed in Ref.~\cite{HFSummaryAarts17}, we will explore the propagation of theoretical uncertainties
in energy-loss predictions, for instance the pp baseline calculation and the (anti-)shadowing parameterization,
in this analysis.

This paper proceeds as follows:
Section~\ref{sec:Method} is dedicated to the introduce the general steps of our hybrid model,
including the initial condition, hydrodynamics expansion of the fireball, heavy quark Brownian motion
and the subsequent hadronization processes.
Section~\ref{sec:Results} presents the results such as the production cross section,
$\raa$ and $\vtwo$ of $D^{+}_{s}$ meson in pp and Pb--Pb collisions.
The comparison with available measurements are performed as well.
Section~\ref{sec:conclusion} contains the summary and conclusion.

%%==============================================
\section{Methodology}\label{sec:Method}
We construct a theoretical framework~\cite{CTGUHybrid1}
to study the charm quark evolution in ultra-relativistic heavy-ion collisions.
The general steps are outlined as follows, as well as the estimation of the theoretical uncertainties.
%%-----------------------------
\subsection{Hybrid Model Construction}\label{subsec:hybrid}
\subsubsection{Initial conditions for the hydrodynamical evolution}
The initial spatial distribution of heavy quark pairs is sampled according to the initial entropy density distributions.
The relevant transverse profile is modeled by a Glauber-based approach~\cite{iEBE},
while the longitudinal profile is described by a data-inspired phenomenological function~\cite{CTGUHybrid1}.
The initial momentum distribution of heavy quark pairs is obtained via the FONLL calculations~\cite{FONLL98, FONLL01, FONLL12}.
Finally, the $c\bar{c}$ is generated in back-to-back before including nuclear shadowing effect~\cite{EPS09}.
%%%%%%%%%%%%%%%%%%%%%%%%%%%%%%%%%%%%%%%%%%%%%%%%%%%%%%%%%

The initial entropy density distributions will be taken as input of the subsequent hydrodynamical evolution,
which can be described by utilizing a 3+1 dimensional relativistic viscous hydrodynamics model~\cite{vhlle}
with the start time scale $\tau_{0}=0.6~{\rm fm}/c$ and the shear viscosity ${\eta/s=1/(4\pi)}$.
The tuning parameters in these modules are determined by the model-to-data comparison~\cite{CTGUHybrid1}.
%%%%%%%%%%%%%%%%%%%%%%%%%%%%%%%%%%%%%%%%%%%%%%%%%%%%%%%%%
\subsubsection{Heavy quark diffusion}\label{subsubsec:HQDiffu}
The Brownian motion of charm quark when propagating through the Quark-Gluon Plasma (QGP), is described by utilizing the Langevin Transport Equation,
and it can be modified to incorporate both the collisional and radiative energy loss processes, which reads~\cite{CaoPRC13}
\begin{equation}\label{eq:TransprotLVEP_Update}
dp=(F^{\rm Drag} + F^{\rm Diff} + F^{\rm Gluon})dt,
\end{equation}
with the drag force 
\begin{equation}\label{eq:DragForce}
F^{\rm Drag}=-\Gamma(p) \cdot p,
\end{equation}
the thermal random force\footnote[5]{Assuming a isotropic momentum-dependence of the diffusion coefficient with the post-point scheme.}
\begin{equation}\label{eq:ThermalForce}
\langle F_{\rm i}^{\rm Diff}(t) \cdot F_{\rm j}^{\rm Diff}(t+n\Delta t)\rangle \equiv \frac{\kappa(p)}{\Delta t}\delta_{\rm ij}\delta_{0n}
\end{equation}
and the recoil force
\begin{equation}\label{eq:RecoilForce}
F^{\rm Gluon}=-\frac{dp^{\rm Gluon}}{dt}.
\end{equation}
$p^{\rm Gluon}$ indicates the momentum of the radiated gluon,
which can be quantified by the pQCD Higher-Twist calculation~\cite{HTPRL04}.
It is assumed~\cite{CaoPRC13} that the fluctuation-dissipation relation
is still validated between the drag (Eq.~\ref{eq:DragForce}) and the diffusion terms (Eq.~\ref{eq:ThermalForce}) in Eq.~\ref{eq:TransprotLVEP_Update}:
\begin{equation}\label{eq:LTEFDR}
\Gamma(p)=\frac{\kappa(p)}{2TE},
\end{equation}
where, $\Gamma(p)$ and $\kappa(p)$ denote the drag and the momentum diffusion coefficients, respectively,
and they can be re-written via the spatial diffusion coefficient $2\pi TD_{s}$~\cite{Moore04},
\begin{equation}\label{eq:Gamma2Ds}
\Gamma=\frac{1}{(2\pi TD_{s})} \cdot \frac{2\pi T^{2}}{E},
\end{equation}
\vspace{-1.5em}
\begin{equation}\label{eq:Kappa2Ds}
\kappa=\frac{1}{(2\pi TD_{s})} \cdot {4\pi T^{3}}.
\end{equation}
Note that the definition of $2\pi TD_{s}$ is extended from zero-momentum to larger momentum region.
As discussed in Ref.~\cite{CTGUHybrid1}, $2\pi TD_{s}$ can be obtained by performing a phenomenological fit analysis with the lattice QCD calculations.
Two approaches are summarized as follows:
\begin{itemize}
  \item \textbf{Model-A}
        \begin{eqnarray}\label{eq:ModelA}
           2\pi TD_{s} = 7
        \end{eqnarray}
In this approach the drag coefficient behaves $\Gamma \propto T^{2}$,
which is similar with the AdS/CFT or pQCD calculation~\cite{Akamatsu09}.
  \item \textbf{Model-B}
    \begin{equation}\label{eq:ModelB} 
      2\pi TD_{s}=1.3 + (\frac{T}{T_{c}})^2
    \end{equation}
    where, $T_{c}$ denotes the critical temperature.
In this approach the drag coefficient behaves a weak $T$-dependence,
which is consistent with the results shown in Ref.~\cite{Das15, GrecoSQM16}.
\end{itemize}

\begin{figure}[!htbp]
\begin{center}
\setlength{\abovecaptionskip}{-0.1mm}
\setlength{\belowcaptionskip}{-1.5em}
\includegraphics[width=.42\textwidth]{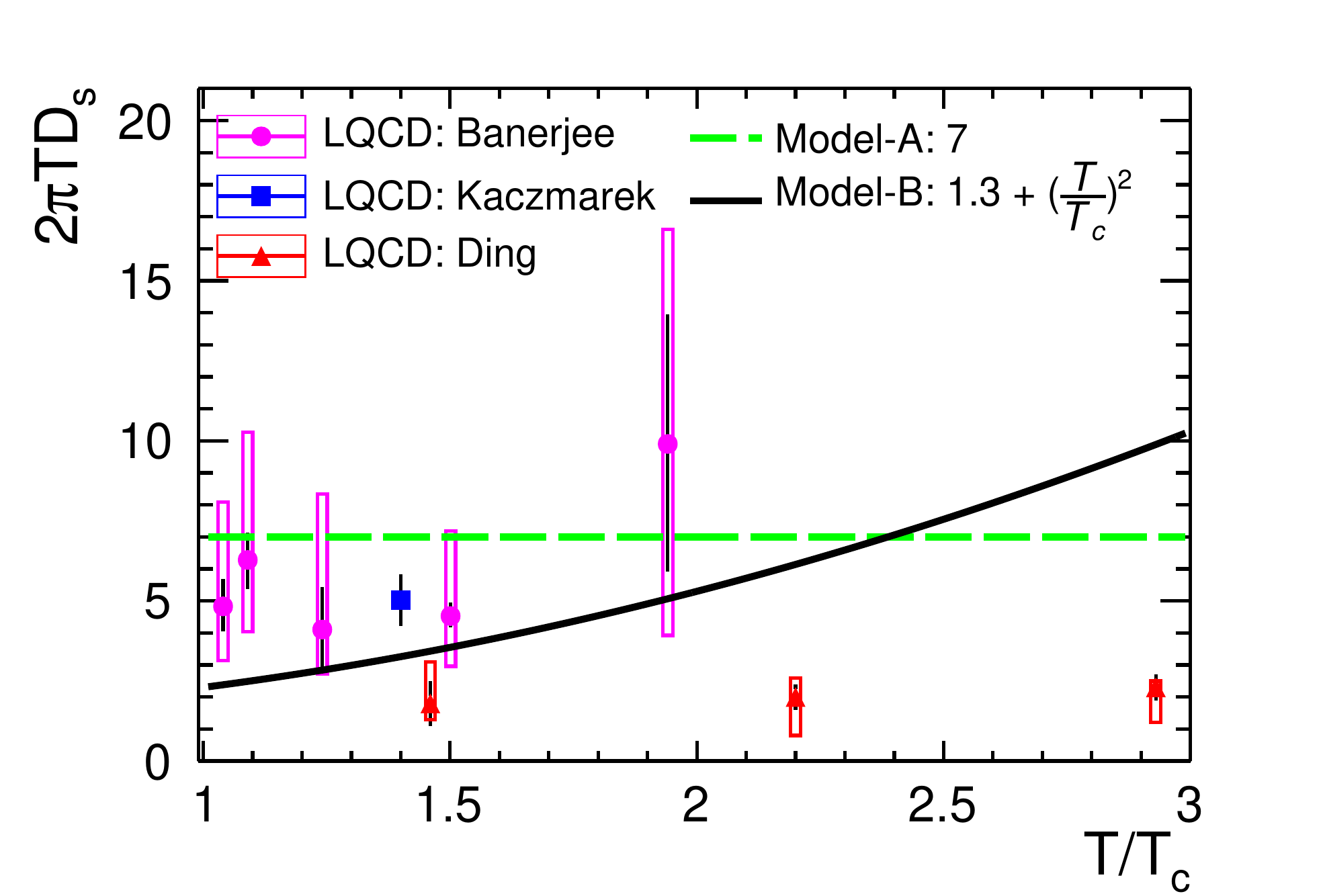}
\caption{(Color online) Charm quark spatial diffusion coefficient $2\pi TD_{s}$
calculated by the lattice QCD at zero momentum: pink circle~\cite{LQCDbanerjee12},
blue square~\cite{LQCDolaf14} and red triangle~\cite{LQCDding12}.
The phenomenological approaches (dashed green and solid black curves) are displayed as well.}
\label{fig:DsVsT}
\end{center}
\end{figure}
Figure~\ref{fig:DsVsT} presents the $T$-dependence of $2\pi TD_{s}$
as calculated by the lattice QCD, i.e. Banerjee (pink circles~\cite{LQCDbanerjee12}),
Kaczmarek (blue square~\cite{LQCDolaf14}) and Ding (red triangle~\cite{LQCDding12}),
as well as the results modeled via the two approaches,
i.e. Model-A (dashed green curve; Eq.~\ref{eq:ModelA}) and Model-B (solid black curve; Eq.~\ref{eq:ModelB}).
The corresponding results are summarized in Tab.~\ref{tab:Coefficients}.
It is found that most of the results obtained for the momentum diffusion coefficient ${\kappa}/{T^{3}}$
and HQ transport coefficient ${\hat{q}_{\rm Q}}/{T^{3}}$,
are consistent with the other model predictions within the significant systematic uncertainties.

\begin{table}[!htbp]
\centering
\begin{tabular}{|c|c|c|c|c|}
\hline
\multicolumn{2}{|c}{\multirow{2}{*}{\centering }}
 & \multicolumn{1}{|c}{\multirow{2}{*}{\centering Model-A}}
 & \multicolumn{1}{|c}{\multirow{2}{*}{\centering Model-B}}
 & \multicolumn{1}{|c|}{\multirow{4}{*}{\centering Reference}}
 \\
\multicolumn{2}{|c}{\centering }
 & \multicolumn{1}{|c}{\centering}
 & \multicolumn{1}{|c}{\centering}
 & \multicolumn{1}{|c|}{}
 \\
\cline{1-4}
\multicolumn{2}{|c}{\multirow{2}{*}{\centering $2\pi TD_{s}$}}
 & \multicolumn{1}{|c}{\multirow{2}{*}{\centering 7}}
 & \multicolumn{1}{|c}{\multirow{2}{*}{\centering $1.3+(\frac{T}{T_{c}})^2$}}
 & \multicolumn{1}{|c|}{\multirow{2}{*}{}}
  \\
\multicolumn{2}{|c}{}
 & \multicolumn{1}{|c}{}
 & \multicolumn{1}{|c}{}
 & \multicolumn{1}{|c|}{}
  \\
\hline
\multicolumn{2}{|c}{\multirow{2}{*}{\centering $\frac{\kappa}{T^{3}}(\frac{T}{T_{c}}=1.5)$}}
 & \multicolumn{1}{|c}{\multirow{2}{*}{\centering 1.80}}
 & \multicolumn{1}{|c}{\multirow{2}{*}{\centering 3.53}}
 & \multicolumn{1}{|c|}{\multirow{2}{*}{\centering $1.8\sim3.4$~\cite{LQCDFrancis15}}}
  \\
\multicolumn{2}{|c}{}
 & \multicolumn{1}{|c}{}
 & \multicolumn{1}{|c}{}
 & \multicolumn{1}{|c|}{}
  \\
\hline
\multicolumn{2}{|c}{\multirow{2}{*}{\centering $\frac{\hat{q}_{\rm Q}}{T^{3}}(\frac{T}{T_{c}}=1.88)$}}
 & \multicolumn{1}{|c}{\multirow{2}{*}{\centering 3.59}}
 & \multicolumn{1}{|c}{\multirow{2}{*}{\centering 5.20}}
 & \multicolumn{1}{|c|}{\multirow{2}{*}{\centering $3.4\sim5.8$~\cite{JETCoef14}}}
  \\
\multicolumn{2}{|c}{}
 & \multicolumn{1}{|c}{}
 & \multicolumn{1}{|c}{}
 & \multicolumn{1}{|c|}{}
  \\
\hline
\multicolumn{2}{|c}{\multirow{2}{*}{\centering $\frac{\hat{q}_{\rm Q}}{T^{3}}(\frac{T}{T_{c}}=2.61)$}}
 & \multicolumn{1}{|c}{\multirow{2}{*}{\centering 3.59}}
 & \multicolumn{1}{|c}{\multirow{2}{*}{\centering 3.11}}
 & \multicolumn{1}{|c|}{\multirow{2}{*}{\centering $2.3\sim5.1$~\cite{JETCoef14}}}
  \\
\multicolumn{2}{|c}{}
 & \multicolumn{1}{|c}{}
 & \multicolumn{1}{|c}{}
 & \multicolumn{1}{|c|}{}
  \\
\hline
\end{tabular}
\caption{Summary of the different approaches for $2\pi TD_{s}$ as a function of temperature (see Fig.~\ref{fig:DsVsT}),
as well as the relevant results obtained for ${\kappa}/{T^{3}}$ and ${\hat{q}_{\rm Q}}/{T^{3}}$.
The other model predictions are shown for comparison.}
\label{tab:Coefficients}
\end{table}

For charm quark, the relevant thermalization time defined in zero momentum limit~\cite{Moore04}
\begin{equation}\label{eq:TauThermal}
\tau_{charm}=\frac{m_{charm}}{2\pi T^{2}_{c}} \cdot \frac{2\pi TD_{s}}{(T/T_{c})^{2}},
\end{equation}
is 3.03 and 2.29 ${\rm fm/{\it c}}$ for Model-A and Model-B, respectively,
in Pb--Pb collisions at 2.76 and 5.02 TeV with $T=2T_{c}=330~{\rm MeV}$ and $m_{charm}=1.5~{\rm GeV}$.

\begin{figure}[!tbp]
\begin{center}
\setlength{\abovecaptionskip}{-0.1mm}
\setlength{\belowcaptionskip}{-1.5em}
\includegraphics[width=.42\textwidth]{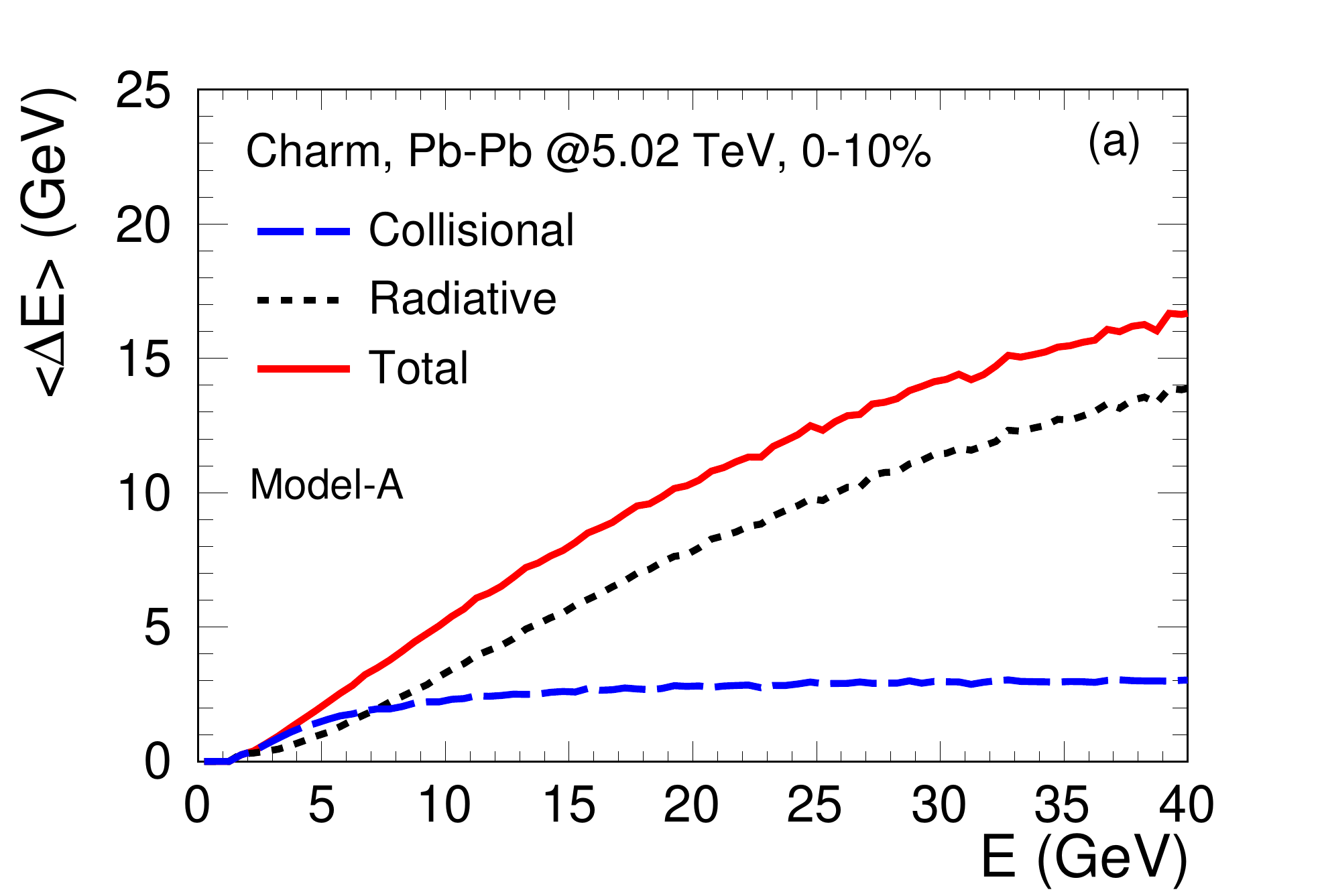}
\includegraphics[width=.42\textwidth]{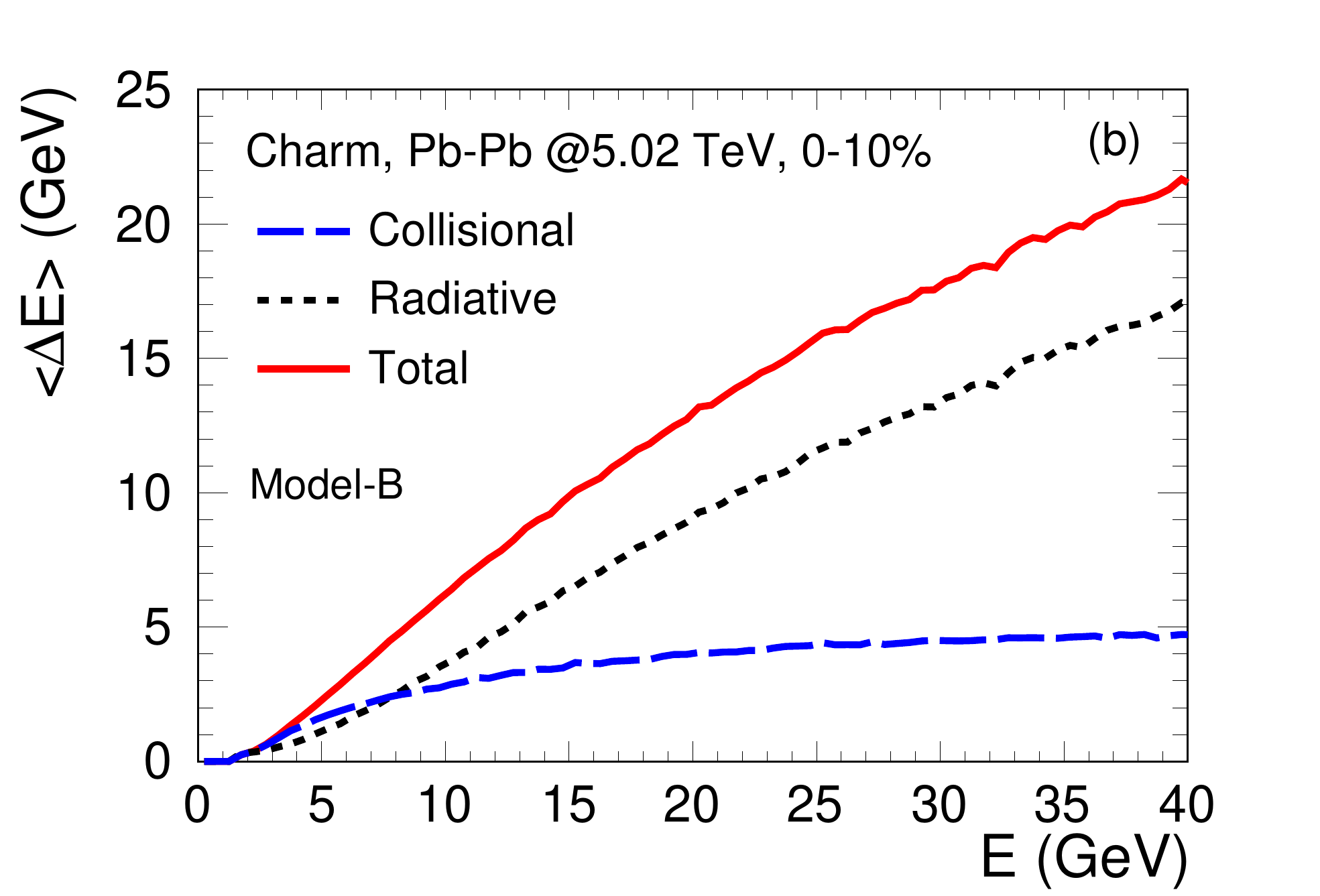}
\caption{(Color online) Energy loss of charm quarks obtained via (a) Model-A
and (b) Model-B: collisional and radiative contributions are shown separately
as long dashed blue and dashed black curves,
respectively, in each panel.
The combined results are shown as solid red curve.}
\label{fig:ElossPbPb5020Cent1ModelAB}
\end{center}
\end{figure}

Figure~\ref{fig:ElossPbPb5020Cent1ModelAB} shows the average in-medium energy loss of charm quarks
as a function of the initial energy in central ($0-10\%$) Pb--Pb collisions at $\snn=5.02~{\rm TeV}$,
displaying separately the contributions of collisional (long dashed blue curve) and radiative (dashed black curve) mechanisms.
The results based on Model-A (Eq.~\ref{eq:ModelA}) are shown in the panel-a (upper).
As pointed in Ref.~\cite{CTGUHybrid1}, the collisional energy loss is significant at low energy,
while radiative energy loss is the dominant mechanism at high energy.
The crossing point between collisional and radiative contributions is around ${\rm E}=7\sim8~{\rm GeV}$.
Since the drag force is proportional to charm velocity $v=p/E$ (Eq.~\ref{eq:DragForce} and \ref{eq:Gamma2Ds}),
in the low energy region (${\rm E}\lesssim7\sim8~{\rm GeV}$), where the relativistic effect is trivial ($E\propto v^{2}$),
the collisional energy loss will be significant.
However, in the very large energy region ($E\gtrsim 20~{\rm GeV}$),
where the untra-relativistic effect should be taken into account ($E\propto 1/\sqrt{1-v^{2}}$ and $v\lesssim 1$),
therefore, the collisional contribution will increase slowly at larger energy.
In this case, the radiative energy loss will be the most dominant energy loss mechanisms.
The results based on Model-B (Eq.~\ref{eq:ModelB}) are displayed in the panel-b (bottom) of Fig.~\ref{fig:ElossPbPb5020Cent1ModelAB}.
A qualitatively similar trend can be found with Model-B, but with slightly stronger energy loss effects.
This is caused by the fact that~\cite{CTGUHybrid1},
$(1)$ the underlying medium temperature drops rapidly from its initial value,
and charm quark will stay longer at low temperature ($\sim 1-2T_{c}$);
$(2)$ the initial transverse momentum spectrum of charm quark is much more harder than that of medium constituent,
thus, the multiple elastic scatterings among them are dominated by the drag term rather than the diffusion term;
$(3)$ a larger drag coefficient near $T_{c}$ with Model-B,
stating a stronger interaction strength between the injected charm quark and the incident medium constituents,
consequently, the charm quark allows to lose more its energy with Model-B approach.

%%%%%%%%%%%%%%%%%%%%%%%%%%%%%%%%%%%%%%%%%%%%%%%%%%%%%%%%%
\subsubsection{Heavy quark hadronization}\label{subsubsec:HQHad}
When the local temperature below the critical one $T_{c}=165~{\rm GeV}$,
the charm quark will undergo the instantaneous hadronization
via a ``dual" approach, including fragmentation and heavy-light coalescence mechanisms.
Concerning the universal fragmentation functions, various models are adopted in this work,
e.g. Lund-PYTHIA 6.4~\cite{PYTHIA64}, Peterson~\cite{FragOriginalPeterson83},
Collins-Spiller~\cite{FragCollJPG}, Braaten~\cite{FragBraaten93} and FONLL-style~\cite{FragFONLLPRL},
which are summarized in Tab.~\ref{tab:FragFuns}.
Apart from the Lund-PYTHIA, the fragmentation fractions for the various hadron species are
$f(c\rightarrow D^{0})=0.566$, $f(c\rightarrow D^{+})=0.227$, $f(c\rightarrow D^{\ast +})=0.230$
and $f(c\rightarrow D_{s}^{+})=0.081$~\cite{CTGUHybrid1}, respectively, in the other approaches.
\begin{table}[!htbp]
\centering
\begin{tabular}{|c|c|c|}
\hline
\multicolumn{1}{|c}{\multirow{2}{*}{\centering Name}}
 & \multicolumn{1}{|c}{\multirow{2}{*}{\centering Frag. Function}}
 & \multicolumn{1}{|c|}{\multirow{2}{*}{\centering Parameter}}
 \\
\multicolumn{1}{|c}{}
 & \multicolumn{1}{|c}{}
 & \multicolumn{1}{|c|}{}
  \\
\hline
\multicolumn{1}{|c}{\multirow{2}{*}{\centering Lund-PYTHIA}}
 & \multicolumn{1}{|c}{\multirow{2}{*}{\centering $\rm Eq.~12.11$ in Ref.~\cite{PYTHIA64}}}
 & \multicolumn{1}{|c|}{\multirow{2}{*}{\centering Default ones}}
  \\
\multicolumn{1}{|c}{}
 & \multicolumn{1}{|c}{}
 & \multicolumn{1}{|c|}{}
  \\
\hline
\multicolumn{1}{|c}{\multirow{2}{*}{\centering Peterson}}
 & \multicolumn{1}{|c}{\multirow{2}{*}{\centering $\rm Eq.~4$ in Ref.~\cite{FragOriginalPeterson83}}}
 & \multicolumn{1}{|c|}{\multirow{2}{*}{\centering $\epsilon_{\rm c}=0.06$}}
  \\
\multicolumn{1}{|c}{}
 & \multicolumn{1}{|c}{}
 & \multicolumn{1}{|c|}{}
  \\
\hline
\multicolumn{1}{|c}{\multirow{2}{*}{\centering Collins-Spiller}}
 & \multicolumn{1}{|c}{\multirow{2}{*}{\centering $\rm Eq.~20$ in Ref.~\cite{FragCollJPG}}}
 & \multicolumn{1}{|c|}{\multirow{2}{*}{\centering $\epsilon_{\rm c}=0.01$}}
  \\
\multicolumn{1}{|c}{}
 & \multicolumn{1}{|c}{}
 & \multicolumn{1}{|c|}{}
  \\
\hline
\multicolumn{1}{|c}{\multirow{2}{*}{\centering Braaten}}
 & \multicolumn{1}{|c}{\multirow{2}{*}{\centering $\rm Eq.~9~and~12$ in Ref.~\cite{FragBraaten93}}}
 & \multicolumn{1}{|c|}{\multirow{2}{*}{\centering $r=\frac{m_{hadron}-m_{c}}{m_{hadron}}$}}
  \\
\multicolumn{1}{|c}{}
 & \multicolumn{1}{|c}{}
 & \multicolumn{1}{|c|}{}
  \\
\hline
\multicolumn{1}{|c}{\multirow{2}{*}{\centering FONLL-style}}
 & \multicolumn{1}{|c}{\multirow{2}{*}{\centering $\rm Eq.~9~and~12$ in Ref.~\cite{FragBraaten93}}}
 & \multicolumn{1}{|c|}{\multirow{2}{*}{\centering $r=0.1$~\cite{FragFONLLPRL}}}
  \\
\multicolumn{1}{|c}{}
 & \multicolumn{1}{|c}{}
 & \multicolumn{1}{|c|}{}
  \\
\hline
\end{tabular}
\caption{Summary of the different fragmentation functions, as well as the relevant parameters.
Note that $m_{hadron}$ and $m_{c}$ in Braaten are the mass of the open charmed hadron
and its mother charm quark, respectively.}
\label{tab:FragFuns}
\end{table}

According to the heavy-light coalescence model~\cite{NewCoal16},
the momentum distributions of heavy-flavor mesons ($Q\bar{q}$) are given as
\begin{equation}
\begin{aligned}\label{eq:MesonCoal}
\frac{dN_{\rm M}}{d^{3}\vec{p}_{\rm M}}=&g_{\rm M}\int d^{3}\vec{x}_{\rm Q}d^{3}\vec{p}_{\rm Q} d^{3}\vec{x}_{\rm\bar{q}}d^{3}\vec{p}_{\rm\bar{q}} f_{\rm Q}(\vec{x}_{\rm Q},\vec{p}_{\rm Q}) f_{\rm\bar{q}}(\vec{x}_{\rm\bar{q}},\vec{p}_{\rm\bar{q}}) \\
&{\overline W}_{\rm M}^{\rm (n)}(\vec{y},\vec{k}) \delta^{(3)}(\vec{p}_{\rm M}-\vec{p}_{\rm Q}-\vec{p}_{\rm\bar{q}})
\end{aligned}
\end{equation}
where, $g_{\rm M}$ is the degeneracy factor;
$f_{\rm Q}(\vec{x}_{\rm Q},\vec{p}_{\rm Q})$ and $f_{\rm\bar{q}}(\vec{x}_{\rm\bar{q}},\vec{p}_{\rm\bar{q}})$
are the phase-space distributions of heavy quark and light anti-quark (i.e. the coalescence candidate), respectively.
The coalescence probability for $Q\bar{q}$ combination to form the heavy-flavor meson in the $n^{th}$ excited state,
is quantified by
\begin{equation}\label{eq:InteWigner2}
{\overline W}_{\rm M}^{\rm (n)}(\vec{y},\vec{k})=\frac{\upsilon^{n}}{n!} e^{-\upsilon}, \qquad
\upsilon=\frac{1}{2}\biggr(\frac{\vec{y}^{\;2}}{\sigma_{\rm M}^{2}}+\sigma_{\rm M}^{2}\vec{k}^{\;2}\biggr),
\end{equation}
where,
\begin{equation}
\begin{aligned}\label{eq:RelativeXP}
&\vec{y}=\vec{x}_{\rm Q}-\vec{x}_{\rm\bar{q}} \\
&\vec{k}=(m_{\rm\bar{q}}\vec{p}_{\rm Q}-m_{\rm Q}\vec{p}_{\rm\bar{q}})/(m_{\rm Q}+m_{\rm\bar{q}})
\end{aligned}
\end{equation}
are the relative coordinate and the relative momentum, respectively, in the center-of-mass frame of $Q\bar{q}$ pair.
The width parameter $\sigma_{\rm M}$ can be written as~\cite{CTGUHybrid1} 
\begin{eqnarray}\label{eq:SigM}
\sigma_{\rm M}^{2}~&&= \left\{ \begin{array}{ll}
\frac{2}{3} \frac{(e_{\rm Q}+e_{\rm\bar{q}})(m_{\rm Q}+m_{\rm\bar{q}})^{2}}{e_{\rm Q}m_{\rm\bar{q}}^{2} + e_{\rm\bar{q}}m_{\rm Q}^{2}} \cdot \langle r_{\rm M}^{2} \rangle & \textrm{\qquad (n=0)} \\
\\
\frac{2}{5} \frac{(e_{\rm Q}+e_{\rm\bar{q}})(m_{\rm Q}+m_{\rm\bar{q}})^{2}}{e_{\rm Q}m_{\rm\bar{q}}^{2} + e_{\rm\bar{q}}m_{\rm Q}^{2}} \cdot \langle r_{\rm M}^{2} \rangle & \textrm{\qquad (n=1)}
\end{array} \right.
\end{eqnarray}
where, $\langle r_{\rm M}^{2} \rangle \approx (0.9~{\rm fm})^{2}$
is the mean-square charge radius of D-meson;
$e_{\rm Q}$ and $e_{\rm\bar{q}}$ are the absolute values of the charge of heavy quark and light anti-quark, respectively;
the light (anti-)quark mass takes $m_{\rm u/\bar{u}}=m_{\rm d/\bar{d}}=300~{\rm MeV}$ and $m_{\rm s/\bar{s}}=475~{\rm MeV}$.

\begin{figure}[!htbp]
\begin{center}
\setlength{\abovecaptionskip}{-0.1mm}
\setlength{\belowcaptionskip}{-1.5em}
\includegraphics[width=.42\textwidth]{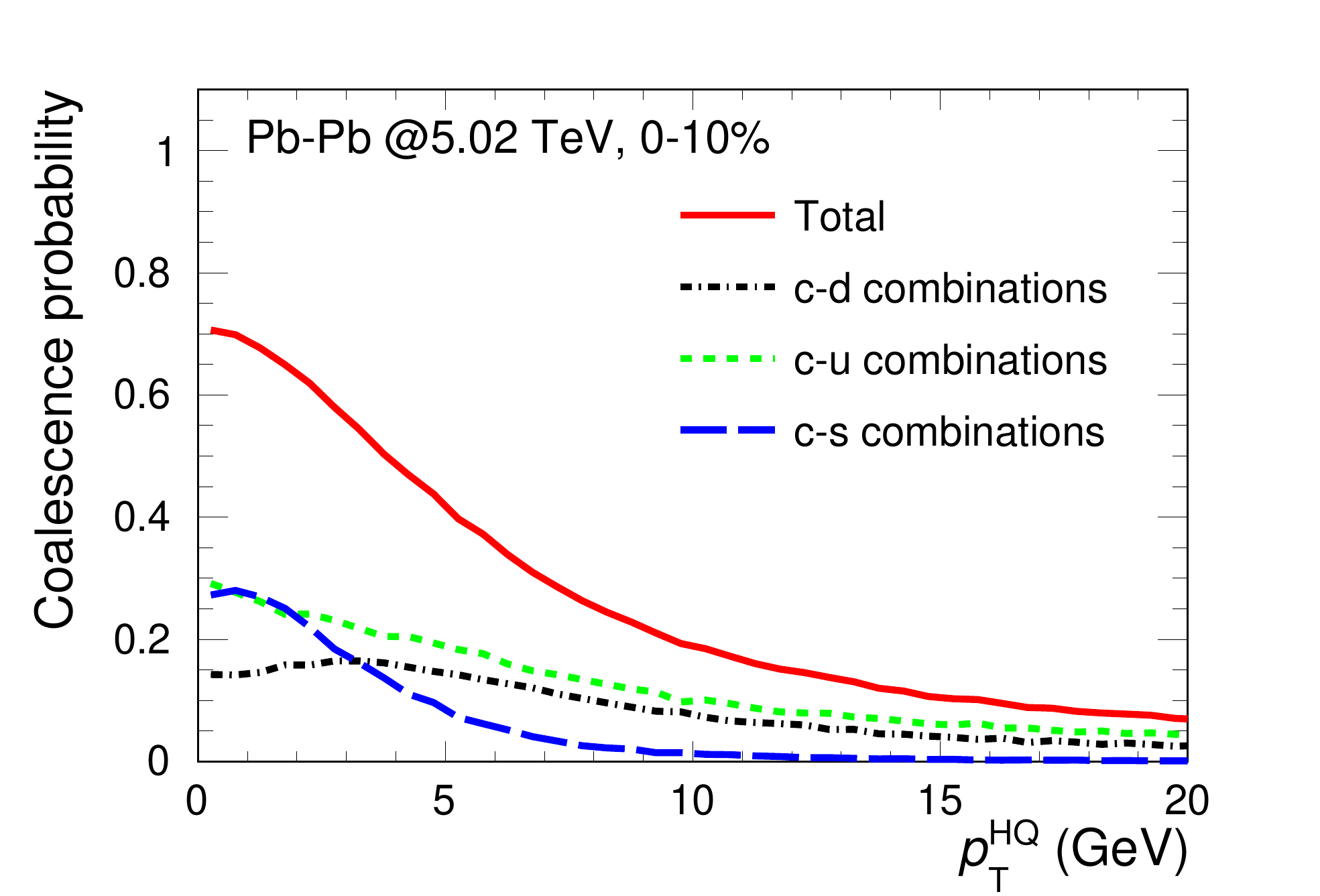}
\caption{(Color online) Comparison of the coalescence probability contributed by different combinations 
in central ($0-10\%$) Pb--Pb collisions at $\snn=5.02~{\rm TeV}$: $cd$ (dot-dashed black curve),
$cu$ (dashed green curve) and $cs$ (long dashed blue curve).
Both the ground states and the first excited states are considered.
The combined results (solid red curve) are presented as well.}
\label{fig:CoalProb}
\end{center}
\end{figure}
We consider the charm-strange meson species up to their first excited states ($n\leqslant1$),
which are listed in details in Ref.~\cite{CTGUHybrid1}.
Figure~\ref{fig:CoalProb} shows the coalescence probability obtained in central ($0-10\%$) Pb--Pb collisions at $\snn=5.02~{\rm TeV}$,
as a function of the charm quark transverse momentum ($\pt$).
The results for the charm quark combined with down quark ($cd$),
up quark ($cu$) and strange quark ($cs$) are presented as dot-dashed black, dashed green and long dashed blue curves, respectively.
As shown in Eq.~\ref{eq:InteWigner2} and~\ref{eq:SigM},
the quark mass and its charge plays the role of the weighting factor in the heavy-light coalescence model,
resulting in the difference among $cd$, $cu$ and $cs$ combinations.
Moreover, this difference can also be induced at a certain amount by the thermal spectrum of $u/d$ and $s$ quark,
which is steeper for the former one, indicating a larger probability to sample the light quark with small $\pt$. 
Finally, it is found that the charm quark prefers to coalesce with $u$ and $s$ quarks in the range $\pt\lesssim3~{\rm GeV}$.
The total results (solid red curve) show a decreasing behavior with increasing $\pt$,
varying from 0.7 at $\pt\sim0$ to 0.2 at $\pt\sim10~{\rm GeV}$,
hence, the HQ with low/moderate and high $\pt$ tends to hadronize via coalescence and fragmentation mechanisms, respectively.

Note that the coalescence candidates are sampled among various light (anti-)quarks, which are assumed to thermalize inside QGP.
Therefore, we utilize the Fermi-Dirac approach, $f_{\rm q}(\vec{p})\propto (e^{\sqrt{{\vec p}^{2} + m_{\rm q}^{2}}/T_{c}} + 1)^{-1}$,
to describe its density distribution, 
where $m_{\rm q}$ is the light (anti-)quark mass, and $T_{c}=165~{\rm MeV}$ is the critical temperature.
The flavor of the light (anti-)quark is determined
according to the integrated parton density $\rho=\int_{-\infty}^{\infty}d^{3}\vec{p}f_{\rm q}(\vec{p})$.
For instance, $\rho_{u/d}=0.18~{\rm fm}^{-3}$ for $u/\bar{u}$ and $d/\bar{d}$ quarks,
and $\rho_{s}=0.10~{\rm fm}^{-3}$ for $s/\bar{s}$ quarks,
resulting in the relative ratio $\rho_{u}:\rho_{d}:\rho_{s}\approx1:1:0.5$,
which is kept during the sampling procedure. 

%%%%%%%%%%%%%%%%%%%%%%%%%%%%%%%%%%%%%%%%%%%%%%%%%%%%%%%%%
\subsection{Theoretical Uncertainty}\label{subsec:systunc}
In this analysis, the total theoretical uncertainty consists of three components:
FONLL predictions, nuclear shadowing and fragmentation models, which are added in quadrature for the final predictions.

The initial charm quark spectra are determined by the FONLL calculations~\cite{FONLL98},
as well as the corresponding central values obtained by setting $\mu_{\rm R}=\mu_{\rm F}=\mu_{0}\equiv\sqrt{\pt^{2}+m_{\rm c}^{2}}$,
where, $\mu_{\rm R}$ ($\mu_{\rm F}$) is the renormalization (factorization) scale;
$m_{\rm c}$ denotes the heavy quark mass, and its central value is $m_{\rm c}=1.5~{\rm GeV}$.
The relevant uncertainties are estimated via a conservative approach~\cite{FONLLErr}:
$\mu_{0}/2<\mu_{\rm R},~\mu_{\rm F}<2\mu_{0}$, $\mu_{\rm R}/2<\mu_{\rm F}<2\mu_{\rm R}$ and $1.3 < m_{c} < 1.7~{\rm GeV}$.

The uncertainty on nuclear shadowing is estimated according to the
various nPDFs sets in EPS09NLO parameterization, which are obtained by tuning the fit parameters to reproduce the available measurements~\cite{EPS09}.
In this work, we employ the nPDFs sets up to $k=7$. See $\rm Eq.~2.12$ and $2.13$ in Ref.~\cite{EPS09} for details.

Based on the different fragmentation scenarios (see Tab.~\ref{tab:FragFuns}),
the final observables such as the production cross section are close to each other.
Therefore, we take the averaged results among them as the final one,
and the maximum dispersion gives the theoretical uncertainty.
%%==============================================
\section{Results}\label{sec:Results}
%%-----------------------------
\subsection{Production cross section in pp collisions}\label{subsec:xsection}
\begin{figure}[!htbp]
\begin{center}
\setlength{\abovecaptionskip}{-0.1mm}
\setlength{\belowcaptionskip}{-1.5em}
\includegraphics[width=.42\textwidth]{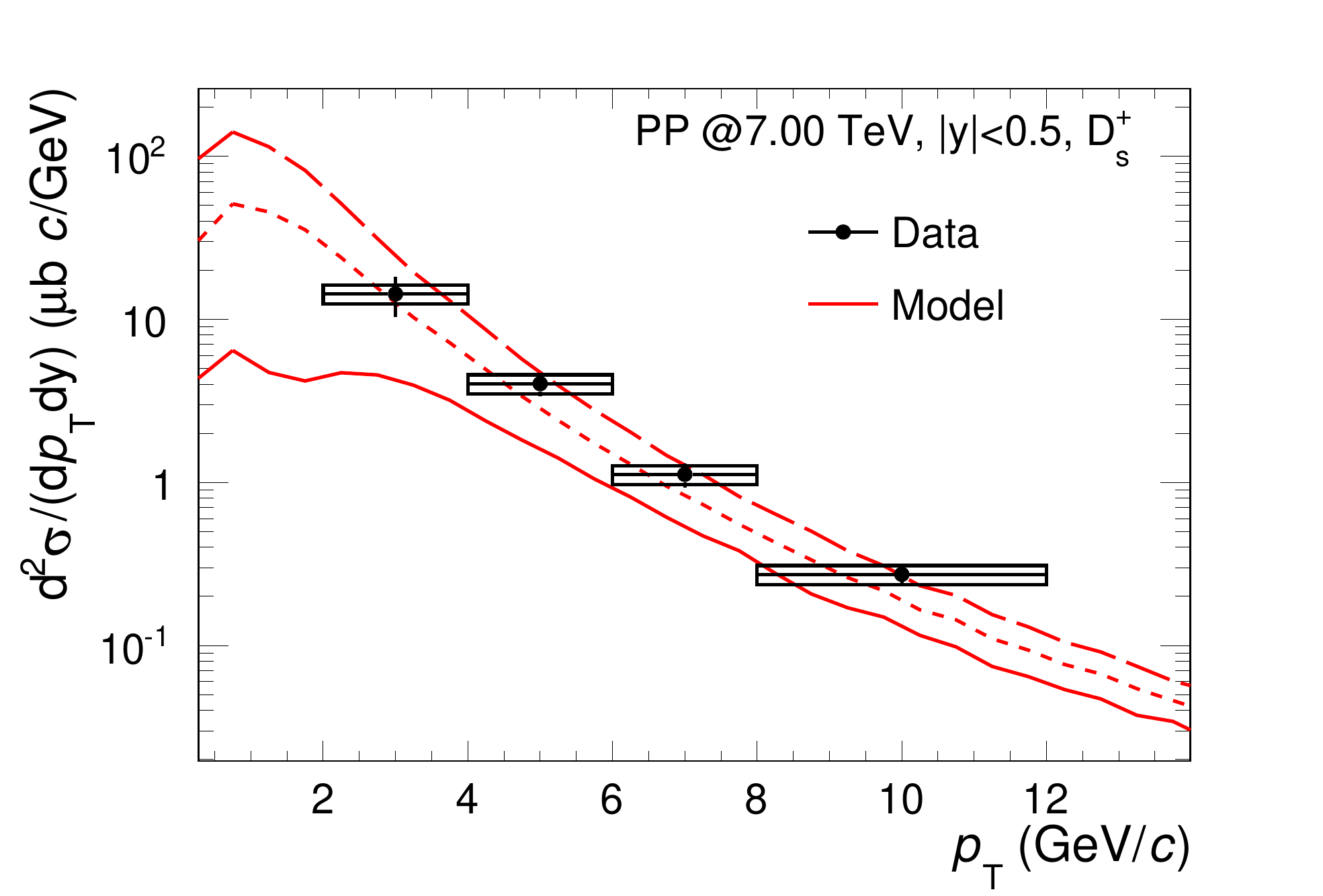}
\caption{(Color online) $\pt$-differential production cross section of $D^{+}_{s}$ meson with $|y|<0.5$ in pp collisions at $\s=7~{\rm TeV}$.
Experimental data taken from Ref.~\cite{ALICEDesonPP7000}.}
\label{fig:SigDs7000}
\end{center}
\end{figure}

In Fig.~\ref{fig:SigDs7000} the $\pt$-differential production cross section of $D^{+}_{s}$ meson
is predicted at mid-rapidity ($|y|<0.5$) in pp collisions at $\s=7~{\rm TeV}$.
The central value, upper band and the lower band are displayed as dashed, long dashed and solid curves, respectively.
The uncertainty on FONLL calculations ($50\sim100\%$) are dominated at $2\lesssim\pt\lesssim8~{\rm GeV/{\it c}}$
comparing with the one on fragmentation models ($\sim30\%$ at maximum),
while they are compatible ($\sim30\%$) toward larger $\pt$.
The experimental data (boxes) are shown for comparison.
Within the experimental and theoretical uncertainties,
the measured $\pt$ dependence can be well described by the model predictions.
Similar conclusion can be found in pp collisions at $\s=5.02~{\rm TeV}.$\footnote[5]{$D^{+}_{s}$ spectrum in pp collisions at $\s=5.02~{\rm TeV}$,
is obtained via $d\sigma_{pp}/d\pt=\raa \cdot d\sigma_{AA}/d\pt$,
while the corresponding $\raa$ and $d\sigma_{AA}/d\pt$ are reported in Ref.~\cite{ALICEDmesonPbPb5020}.}

\begin{figure}[!htbp]
\begin{center}
\setlength{\abovecaptionskip}{-0.1mm}
\setlength{\belowcaptionskip}{-1.5em}
\includegraphics[width=.42\textwidth]{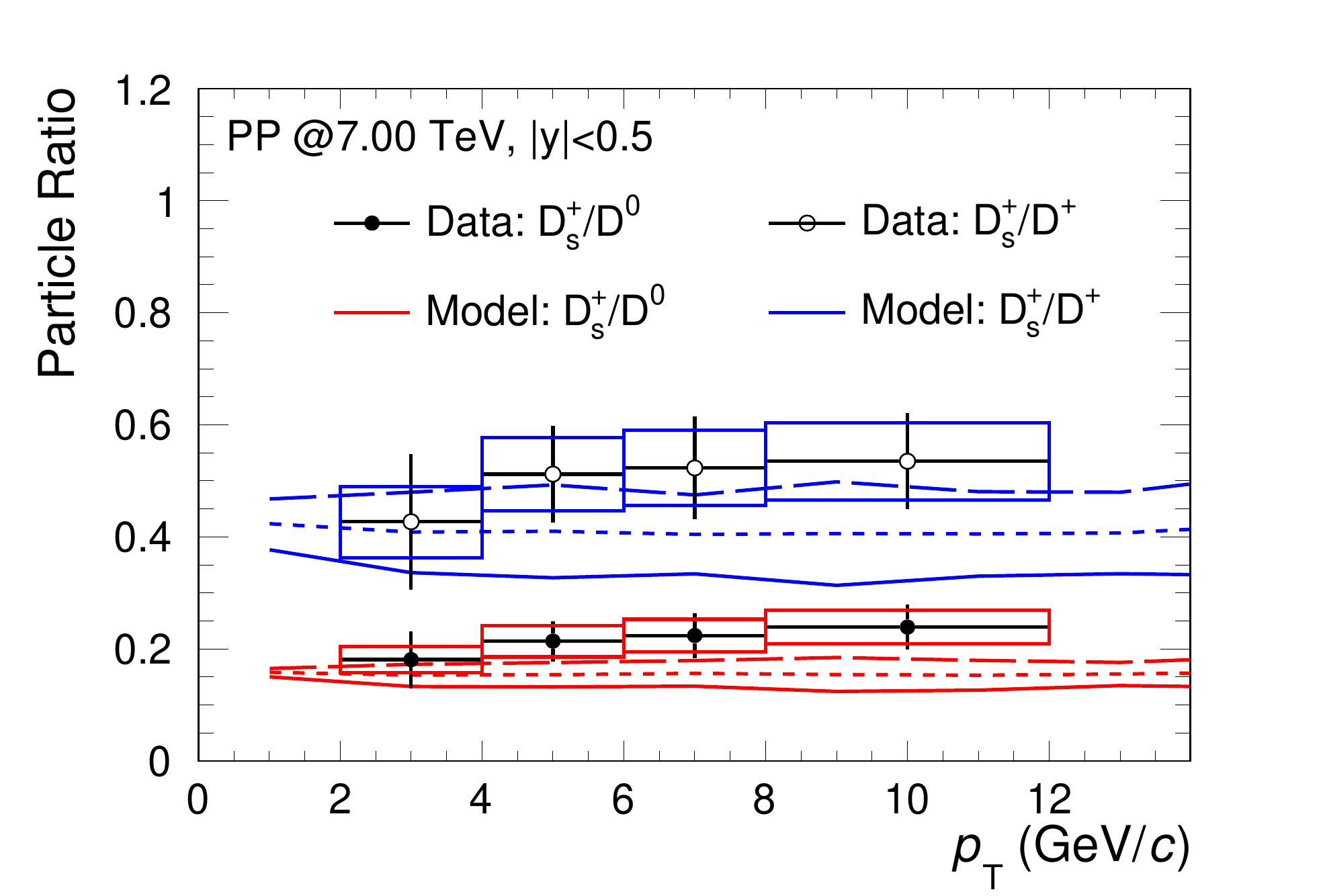}
\caption{(Color online) Ratios of D-meson production cross sections as a function of $\pt$.
the measurements for $D^{+}_{s}/D^{0}$ (solid) and $D^{+}_{s}/D^{+}$ (empty) are shown as boxes,
while the relevant model predictions displayed as the bands.
Experimental data taken from Ref.~\cite{ALICEDesonPP7000}.}
\label{fig:PartRatio7000}
\end{center}
\end{figure}
Figure~\ref{fig:PartRatio7000} presents the ratios of
the charm-strange meson $D^{+}_{s}$ with respect to the non-strange charmed meson such as $D^{0}$ and $D^{+}$,
in pp collisions at $\s=7~{\rm TeV}$. See the legend for details.
For both $D^{+}_{s}/D^{0}$ and $D^{+}_{s}/D^{+}$,
the theoretical uncertainty on FONLL calculation ($\sim10\%$ at maximum) is dominated in the range $2\lesssim\pt\lesssim4~{\rm GeV/{\it {c}}}$,
while the one on fragmentation models ($\sim20\%$ at maximum) at higher $\pt$.
It is found that, within uncertainties, the measurements can be reproduced by the corresponding model predictions.
%%-----------------------------

\begin{figure}[!htbp]
\begin{center}
\setlength{\abovecaptionskip}{-0.1mm}
\setlength{\belowcaptionskip}{-1.5em}
\includegraphics[width=.42\textwidth]{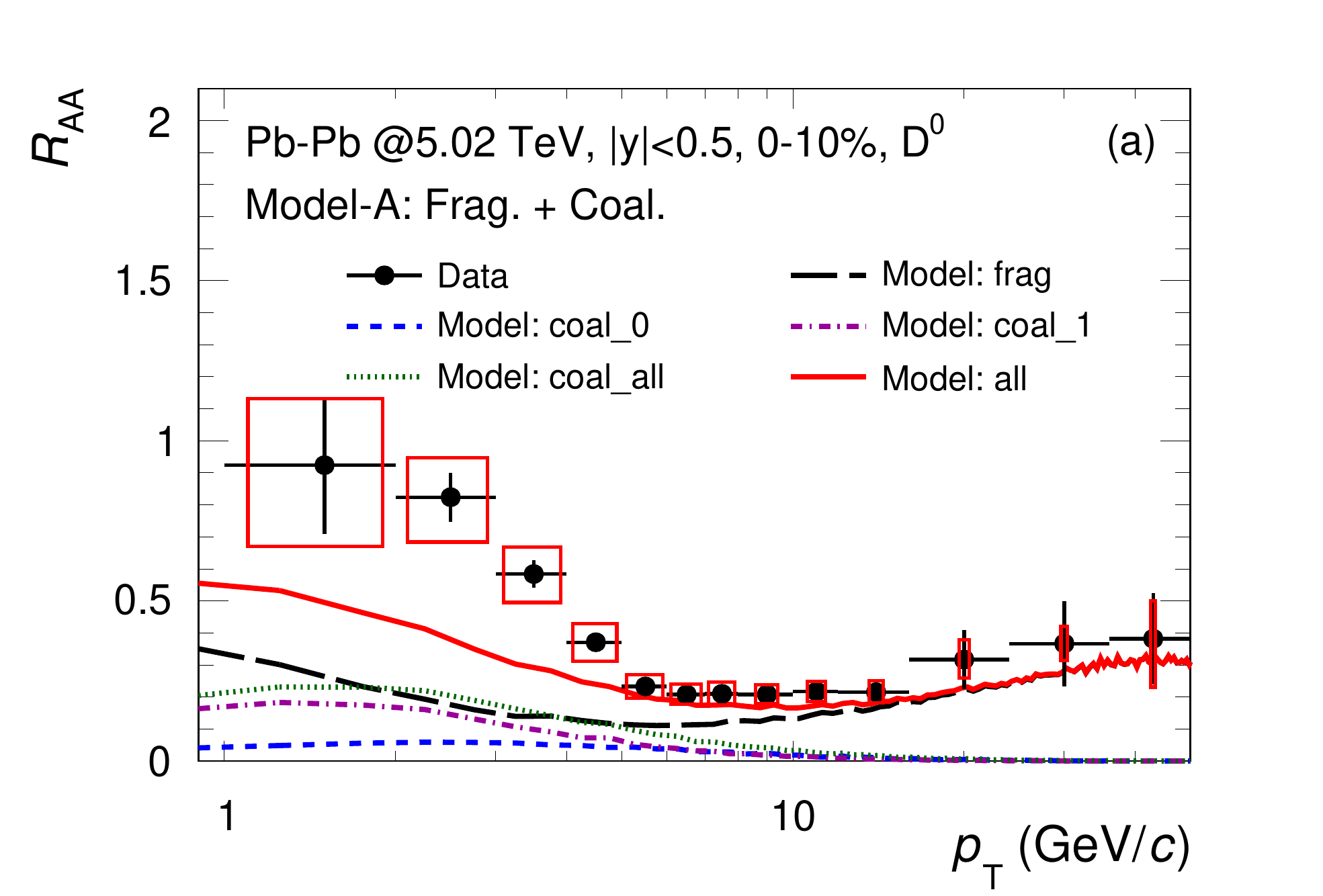}
\includegraphics[width=.42\textwidth]{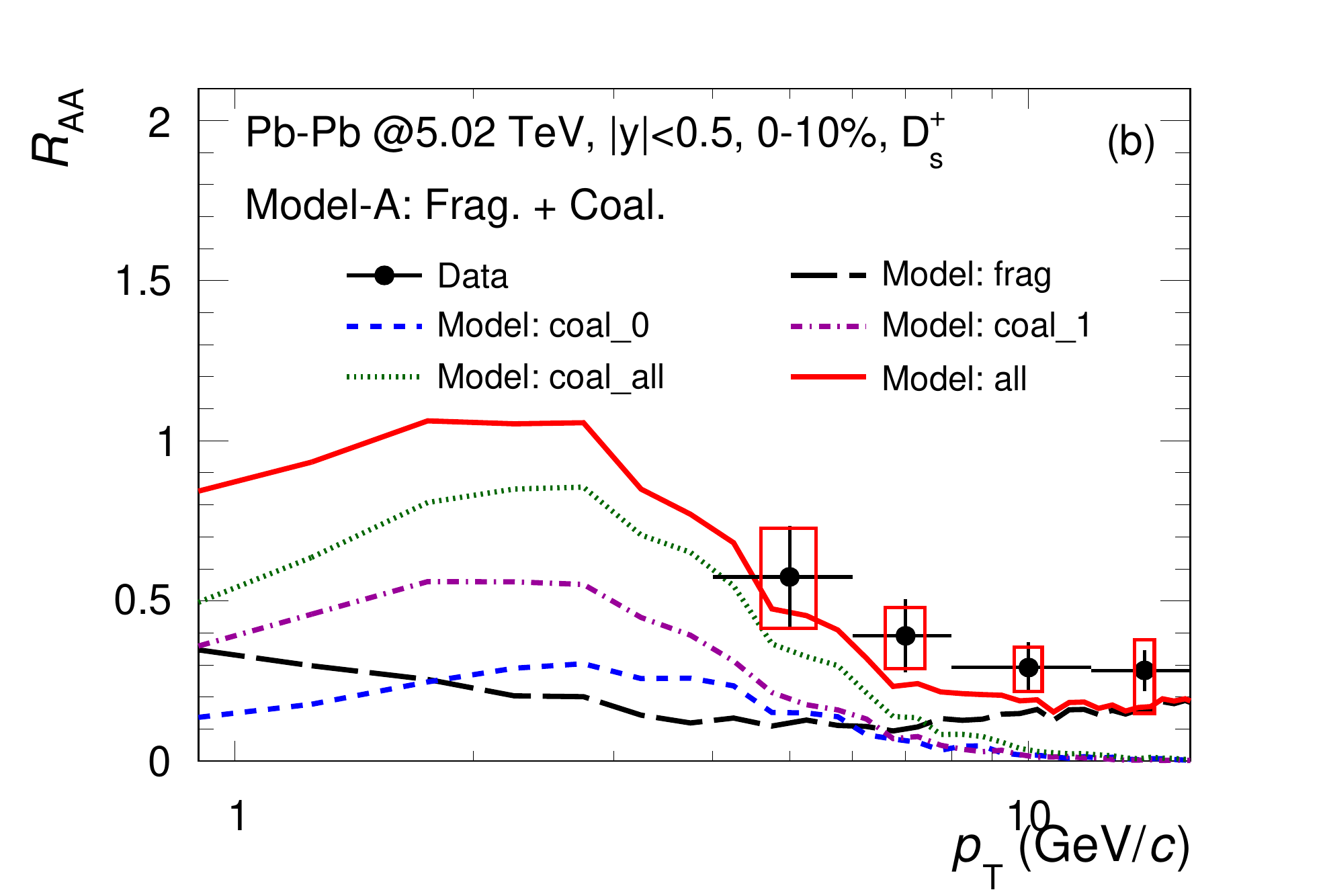}
\caption{(Color online) Comparison of the central predictions of $\raa$ contributed by different hadronization mechanisms,
for (a) the non-strange charmed meson (average among $D^{0}$, $D^{+}$ and $D^{*+}$) and
(b) the charm-strange meson ($D^{+}_{s}$), at mid-rapidity ($|y|<0.5$) in central ($0-10\%$) Pb--Pb collisions at $\snn=5.02~{\rm TeV}$.
See legend for details.  Experimental data taken from Ref.~\cite{ALICEDmesonPbPb5020}.}
\label{fig:CoalEff_5020_c0_c10_ModelA}
\end{center}
\end{figure}

\begin{figure}[!htbp]
\begin{center}
\setlength{\abovecaptionskip}{-0.1mm}
\setlength{\belowcaptionskip}{-1.5em}
\includegraphics[width=.42\textwidth]{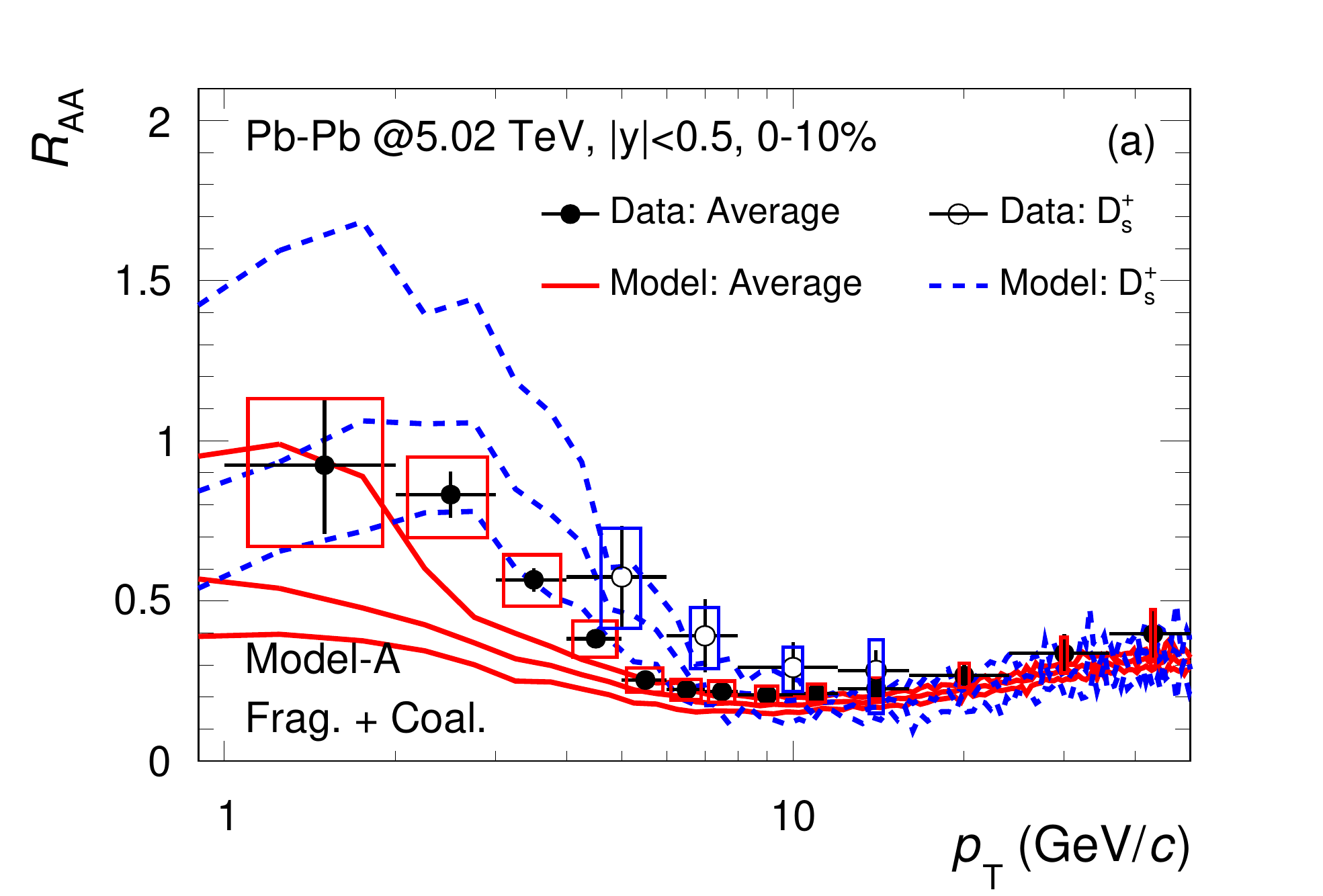}
\includegraphics[width=.42\textwidth]{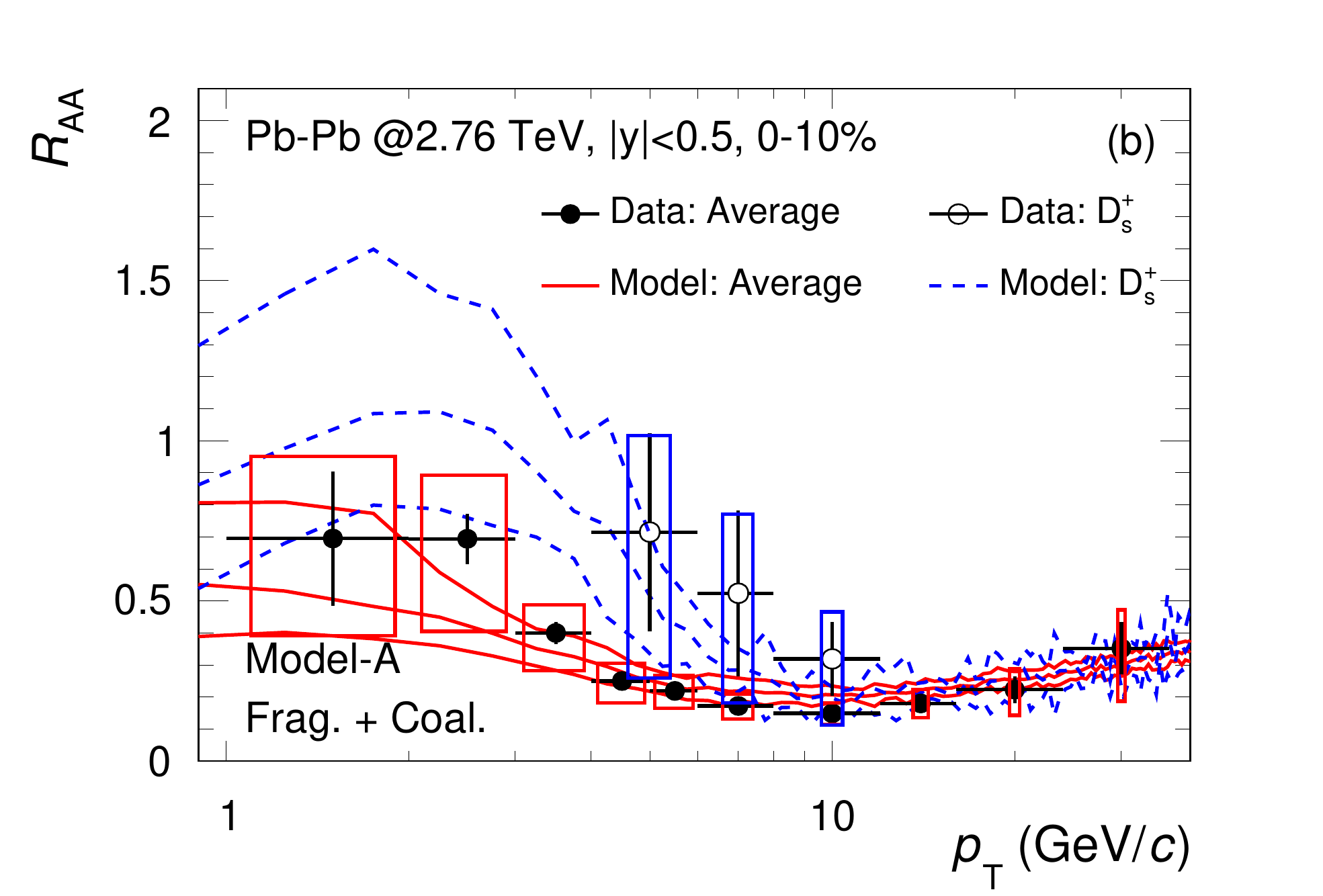}
\caption{(Color online) Comparison of $\raa$ for the average non-strange D mesons (solid red curves) and $D^{+}_{s}$ (dashed blue curves)
predicted at mid-rapidity ($|y|<0.5$) in central ($0-10\%$) Pb--Pb collisions at (a) $\snn=5.02~{\rm TeV}$ and (b) $2.76~{\rm TeV}$, respectively.
Experimental data taken from Ref.~\cite{ALICEDmesonPbPb5020, ALICEDesonPbPb2760RAA}.}
\label{fig:RAA_AllDs_c0_c10_ModelA}
\end{center}
\end{figure}

\subsection{Nuclear modification factor and Elliptic flow}\label{subsec:RAAV2}
The panel-a of Fig.~\ref{fig:CoalEff_5020_c0_c10_ModelA} shows the average $\raa$ of
non-strange charmed meson ($D^{0}$, $D^{+}$ and $D^{*+}$) at mid-rapidity ($|y|<0.5$),
with Model-A approach (Eq.~\ref{eq:ModelA}) in central ($0-10\%$) Pb--Pb collisions at $\snn=5.02~{\rm TeV}$,
which is contributed by the various hadronizaton mechanisms.
It is found that the fragmentation component (long dashed black curve) is dominated at $\pt\gtrsim7-8~{\rm GeV/{\it c}}$,
while the coalescence (dotted green curve) is significant at $1\lesssim\pt\lesssim5~{\rm GeV/{\it c}}$,
and furthermore, the first excited states contribution (dot-dashed purple curve) is more pronounced in this region.
The central prediction (solid red curve) can describe the measurement in the range $\pt>5~{\rm GeV/{\it c}}$.
$\raa$ of charm-strange meson ($D^{+}_{s}$) is presented in the panel-b of Fig.~\ref{fig:CoalEff_5020_c0_c10_ModelA}.
Similar behavior is observed when comparing with the non-strange charmed mesons,
however, the coalescence effect is more pronounced for $D^{+}_{s}$ meson.
It is further checked that $\raa(average)$ and $\raa(D^{+}_{s})$
calculated by considering alone the fragmentation mechanism, are close to each other. 
All the conclusions drawn above are the same as the ones found
in semi-central ($30-50\%$) Pb--Pb collisions at $\snn=5.02~{\rm TeV}$,
as well as in Pb--Pb collisions at $\snn=2.76~{\rm TeV}$.
Therefore, the future measurements of $\raa(D^{+}_{s})$ with higher precision
are more powerful to constrain the heavy-light coalescence effect at moderate $\pt$ ($\pt=2\sim5~{\rm GeV/{\it c}}$).

$\raa(average)$ (solid curves) and $\raa(D^{+}_{s})$ (dashed curves) obtained at mid-rapidity ($|y|<0.5$),
with Model-A approach in central ($0-10\%$) Pb--Pb collisions at $\snn=5.02~{\rm TeV}$,
are shown in the panel-a of Fig.~\ref{fig:RAA_AllDs_c0_c10_ModelA}.
The results between average non-strange D-meson and $D^{+}_{s}$ are similar in the range $\pt\gtrsim6~{\rm GeV/{\it c}}$ ($\pt\gg m_{c}$),
while the latter one is systematically larger at $2<\pt<5~{\rm GeV/{\it c}}$,
resulting in an enhancement of $D^{+}_{s}$ production with respect to the average one.
As mentioned in Fig.~\ref{fig:CoalEff_5020_c0_c10_ModelA},
the enhancement behavior is mainly induced by the coalescence mechanism during the charm quark hadronization.
Note that, for $\raa(average)$, the uncertainty on FONLL calculations ($\sim80\%$ at maximum) are dominated at $1<\pt<3~{\rm GeV/{\it c}}$,
while the one on nuclear shadowing ($\lesssim10\%$) and fragmentation functions ($\sim10-15\%$) are significant at higher $\pt$.
Similar behavior can be found for $\raa(D^{+}_{s})$ at low $\pt$,
but the uncertainty on fragmentation functions ($\sim20-40\%$) are dominated at $\pt>3~{\rm GeV/{\it c}}$.
For comparison, the available measurements for the average non-strange D-meson (solid) and $D^{+}_{s}$ meson (empty) are displayed as well.
Within the experimental and theoretical uncertainties,
the model calculations can reproduce the data for both the average and $D^{+}_{s}$ meson.
Similar results are found in central ($0-10\%$) Pb--Pb collisions at $\snn=2.76~{\rm TeV}$.
See the panel-b of Fig.~\ref{fig:RAA_AllDs_c0_c10_ModelA} for details.

\begin{figure}[!htbp]
\begin{center}
\setlength{\abovecaptionskip}{-0.1mm}
\setlength{\belowcaptionskip}{-1.5em}
\includegraphics[width=.42\textwidth]{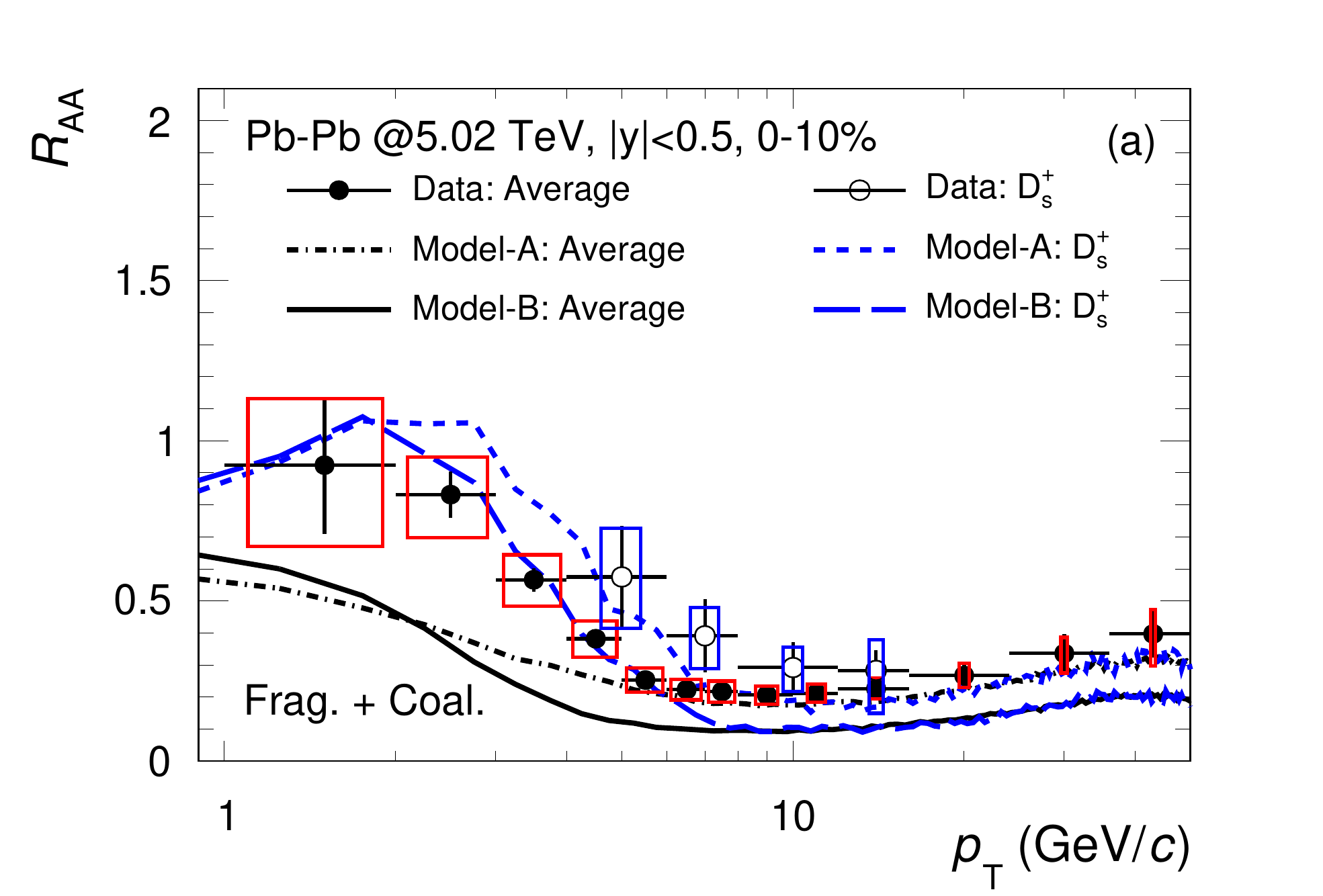}
\includegraphics[width=.42\textwidth]{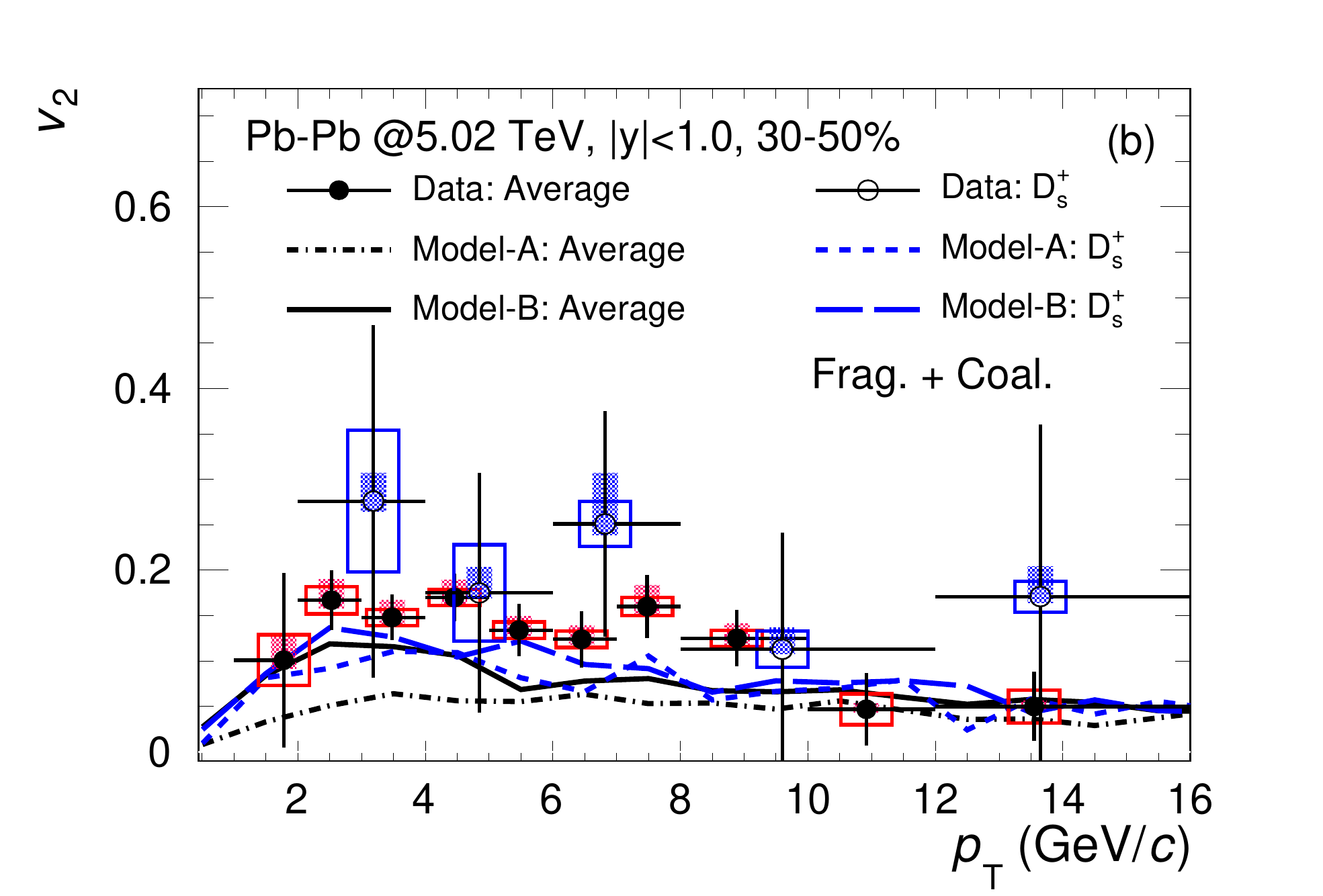}
\caption{Comparison between the average non-strange D mesons and $D^{+}_{s}$ observables
obtained with Model-A and Model-B approaches in Pb--Pb collisions at $\snn=5.02~{\rm TeV}$:
$\raa$ (upper) in $0-10\%$ and $\vtwo$ in $30-50\%$.
Experimental data taken from Ref.~\cite{ALICEDmesonPbPb5020, ALICEDesonPbPb5020V2}.}
\label{fig:RAAV2_PbPb5020_c0_c10_ModelAB}
\end{center}
\end{figure}
To compare the predictions based on Model-A (Eq.~\ref{eq:ModelA}) and Model-B approaches (Eq.~\ref{eq:ModelB}),
the panel-a of Fig.~\ref{fig:RAAV2_PbPb5020_c0_c10_ModelAB} presents $\raa(average)$ and $\raa(D^{+}_{s})$
calculated in central ($0-10\%$) Pb--Pb collisions at $\snn=5.02~{\rm TeV}$.
$\raa(average)$ is enhanced (suppressed) at low (high) $\pt$ from Model-A (dot-dashed black curve) to Model-B (solid black curve),
since the relevant $2\pi TD_{s}$ near $T_{c}$ is smaller based on Model-B, the larger is its initial drag term,
which is more powerful to pull $c\bar{c}$ pairs from high momentum to low momentum~\cite{CTGUHybrid1},
as pointed in Sec.~\ref{subsubsec:HQDiffu}.
The results between different models are close at $\pt\sim2~{\rm GeV/{\it c}}$.
Similar behavior is observed for $\raa(D^{+}_{s})$.
When comparing $\raa(average)$ with $\raa(D^{+}_{s})$,
the results with Model-A approach are discussed already in the panel-a of Fig.~\ref{fig:RAA_AllDs_c0_c10_ModelA},
and same conclusion can be drawn with Model-B approach.
The panel-b of Fig.~\ref{fig:RAAV2_PbPb5020_c0_c10_ModelAB} shows
the elliptic flow coefficient $\vtwo$ predicted in semi-central ($30-50\%$) Pb--Pb collisions at $\snn=5.02~{\rm TeV}$:
both $\vtwo(average)$ and $\vtwo(D^{+}_{s})$ are significantly enhanced
at intermediate $\pt$ ($2\lesssim\pt\lesssim4~{\rm GeV/{\it c}}$) from Model-A (dot-dashed black curve) to Model-B (solid black curve).
Performing the model-to-data comparison,
it is realized that $\raa(average)$ and $\raa(D^{+}_{s})$ favor Model-A to have a better description of the measured $\pt$ dependence,
while their $\vtwo$ prefer Model-B at moderate $\pt$ ($2\lesssim\pt\lesssim4~{\rm GeV/{\it c}}$),
indicating the necessity to consider the temperature- and/or momentum-dependence of $2\pi TD_{s}$ to
describe simultaneously $\raa$ and $\vtwo$ for both the non-strange D-meson and $D^{+}_{s}$ meson.

%%==============================================
\section{Summary and Conclusions}\label{sec:conclusion}
In this analysis, we aim to investigate the nuclear modification of $D^{+}_{s}$ meson spectra
together with its elliptic flow in ultra-relativistic heavy-ion collisions.
We utilize the theoretical framework built in our previous work to achieve this goal,
and extend it to include further the theoretical uncertainty on initial charm quark spectra,
nuclear shadowing and fragmentation model.
The coupling strength for charm quark $2\pi TD_{s}$ is obtained by fitting the lattice QCD calculations:
$2\pi TD_{s}=7$ (\textbf{Model-A}, i.e. no temperature dependence)
and $2\pi TD_{s}=1.3 + (T/T_{c})^2$ (\textbf{Model-B}, i.e. weak temperature dependence).

It is found that $D^{+}_{s}$ spectra measured at mid-rapidity ($|y|<0.5$)
can be well described by the relevant model predictions in pp collisions both at $\s=7~{\rm TeV}$ and $5.02~{\rm TeV}$,
as well as the derived particle ratios $D^{+}_{s}/D^{0}$ and $D^{+}_{s}/D^{+}$.
The nuclear modification factor $\raa(D^{+}_{s})$ is systematically larger than $\raa(non-strange)$
at intermediate $\pt$ ($2\lesssim\pt\lesssim5~{\rm GeV}$) in central ($0-10\%$) and semi-central ($30-50\%$) Pb--Pb collisions
both at $\snn=5.02~{\rm TeV}$ and $2.76~{\rm TeV}$,
which is mainly induced by the heavy-light coalescence mechanism.
Hence, the future measurements of $\raa(D^{+}_{s})$ with higher precision
are more powerful to constrain the heavy-light coalescence effect at moderate $\pt$ ($\pt=2\sim5~{\rm GeV/{\it c}}$).
For the model-to-data comparisons,
the predictions of $\raa(D^{+}_{s})$ and $\raa(non-strange)$ favor Model-A to reproduce well the measured $\pt$ dependence in both colliding energies,
while their $\vtwo$ prefer Model-B at moderate $\pt$ ($2\lesssim\pt\lesssim4~{\rm GeV}$),
suggesting a temperature and/or momentum dependent $2\pi TD_{s}$
is needed to describe simultaneously the D-meson $\raa$ and $\vtwo$ data.
%%==============================================
\begin{acknowledgments}
The authors are grateful to Kejun Wu and Kyong Chol Han for the useful discussions.
S.~Li is supported by the CTGU No.1910103, B2018023, QLPL2018P01, NSFC No.11447023 and No.11875178.
C.~W.~Wang acknowledges the support from the NSFHB No.2012FFA085.
\end{acknowledgments}

%%==============================================
%\bibliography{Shuang_2018}

\end{document}